\documentclass[conference,10pt]{IEEEtran}

\ifCLASSINFOpdf
\else
\fi

\usepackage{color}
\usepackage{amsmath,amssymb,amscd}
\usepackage{subfigure, epsfig}
\usepackage{pstricks}
\usepackage{pstricks-add}
\usepackage{pst-plot}

\newtheorem{theorem}{Theorem}
\newtheorem{definition}[theorem]{Definition}

\newcommand{\bd}{\textbf}

\newcommand{\R}{\mathbb{R}}

\newcommand{\N}{\mathbb{N}}

\newcommand{\C}{\mathcal{C}}

\newcommand{\mc}{\mathcal}

\newgray{darkgray}{.35}
\newgray{cleargray}{.85}

\begin{document}
%
\title{Resilient Source Coding\\
}


\author{\IEEEauthorblockN{Ma\"{e}l Le Treust}
\IEEEauthorblockA{Laboratoire des Signaux et Syst\`{e}mes\\
CNRS - Sup\'{e}lec - Univ. Paris Sud 11\\
91191, Gif-sur-Yvette, France\\
Email: mael.letreust@lss.supelec.fr}
\and
\IEEEauthorblockN{Samson Lasaulce}
\IEEEauthorblockA{Laboratoire des Signaux et Syst\`{e}mes\\
CNRS - Sup\'{e}lec - Univ. Paris Sud 11\\
91191, Gif-sur-Yvette, France\\
Email: samson.lasaulce@lss.supelec.fr}
}


%


\maketitle

\begin{abstract}
This paper provides a source coding theorem for multi-dimensional information signals when, at a given instant, the distribution associated with one arbitrary component of the signal to be compressed is not known and a side information is available at the destination. This new framework appears to be both of information-theoretical and game-theoretical interest~: it provides a new type of constraints to compress an information source; it is useful for designing certain types of mediators in games and characterize utility regions for games with signals. Regarding the latter aspect, we apply the derived source coding theorem to the prisoner's dilemma and the battle of the sexes.
\end{abstract}



%
\IEEEpeerreviewmaketitle

\section{Introduction}\label{introduction}

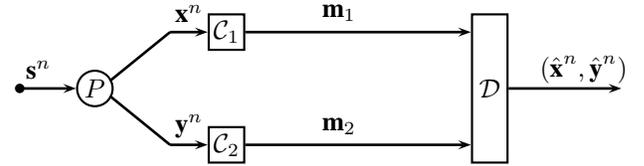
\begin{figure}[!ht]
\begin{center}
\psset{xunit=0.5cm,yunit=0.5cm}
\begin{pspicture}(0,0)(15.5,4.5)
\pscircle(2,2){0.25}
\psframe(5,0)(6,1)
\psframe(5,3)(6,4)
\psframe(12,0)(13,4)
\psdots(0,2)
\psline[linewidth=1pt]{->}(0,2)(1.5,2)
\psline[linewidth=1pt](2.4,2.2)(4,3.5)
\psline[linewidth=1pt](2.4,1.8)(4,0.5)
\psline[linewidth=1pt]{->}(4,3.5)(5,3.5)
\psline[linewidth=1pt]{->}(4,0.5)(5,0.5)
\psline[linewidth=1pt]{->}(6,3.5)(12,3.5)
\psline[linewidth=1pt]{->}(6,0.5)(12,0.5)
\psline[linewidth=1pt]{->}(13,2)(16,2)
\rput[u](0.5,2.5){$\bd{s}^n$}
\rput(2,2){$P$}
\rput(5.5,3.5){$\mc{C}_1$}
\rput(5.5,0.5){$\mc{C}_2$}
\rput(12.5,2){$\mc{D}$}
\rput[u](4.5,4){$\bd{x}^n$}
\rput[u](4.5,1){$\bd{y}^n$}
\rput[u](8.5,4){$\bd{m}_1$}
\rput[u](8.5,1){$\bd{m}_2$}
\rput[u](15,2.5){$(\hat{\bd{x}}^n,\hat{\bd{y}}^n)$}
\end{pspicture}
\caption{The pair of source symbols $(\bd{x}, \bd{y}) \in \mc{X}\times \mc{Y}$ is drawn from the transition probability $P(\bd{x}, \bd{y}|s)$ which depend on a state parameter $s\in \mc{S}$. This parameter varies from stage to stages in an arbitrarily manner. The encoder $\mc{C}_1$ takes a sequence of symbols $\bd{x}^n$ and encode it into a message $\bd{m}_1$.  The encoder $\mc{C}_2$ takes a sequence of symbols $\bd{y}^n$ and encode it into a message $\bd{m}_2$. The decoder observes the pair of messages $\bd{m}_1, \bd{m}_2$ and decode a pair of sequence of symbols $(\hat{\bd{x}}^n,\hat{\bd{y}}^n)$.}\label{fig:AVCSmodel}
\end{center}
\end{figure}

Shannon has shown \cite{shannon-bell-1948} that it is possible to describe without any loss a discrete random variable $\bd{x} \in \mathcal{X}$ (representing an information source) provided that $H(\bd{x})$ bits per information source sample are used, $H(\bd{x})$ being the entropy of the random variable $\bd{x}$. This result holds for a stationary source that is, the distribution of $\bd{x}$, say $P(\cdot)\in \Delta(\mc{X})$, does not vary over time. In \cite{Dobrusin(Indiv)63,Dobrusin(Unif)63}, Dobru\v{s}in introduced a generalized version of this problem where the source distribution can vary. In such a situation, the source distribution $P(\cdot|s)\in \Delta(\mc{X})$ can vary from sample to sample, depending on the parameter or state $s \in \mc{S}$. This problem, which is referred to as the problem of arbitrary varying sources, has been solved in \cite{Berger(Game)71}. In \cite{Gallager(AVS)76} the author solve the problem of arbitrary varying sources when the sequence of
 states $s^n = (s_1, s_2, ...,s_n)$ is known to the destination (i.e., the decoder). The problem of arbitrarily varying source (AVS) is different to the one of universal source coding (USC), even if in both situations the encoder and decoder do not know the right distribution $P(\cdot)\in \Delta(\mc{X})$ of $\bd{x}$. Indeed, the distribution probability of the symbols is constant for the USC problem but vary from stage to stage for the AVS problem. A further generalization of this work has been done in \cite{Ahlswede(ColoringPart1)79} and \cite{Ahlswede(ColoringPart2)80}. Indeed, the latter references deal with the scenario of two correlated sources either in the case where the destination is informed with the sequence of states or the case where it is not known. The work reported in this paper is precisely related to the scenario of two arbitrary varying correlated sources with an uninformed destination; this scenario is described by Fig. \ref{fig:AVCSmodel}. For this scenario, which is motivated next, the achievable rate region in \cite{Ahlswede(ColoringPart1)79} is characterized for the family of information sources satisfying an entropy positiveness condition:
\begin{equation}
\label{eq:entropy-positiveness}
\forall s\in \mc{S}, \ \ H(\bd{x}(s)|\bd{y}(s)) \times H(\bd{y}(s)|\bd{x}(s)) >0,
\end{equation}
where $\mc{S}$ is the set of possible states, $\bd{x}(s)$ and $\bd{y}(s)$ represent the two information sources to be encoded when the state is $s\in \mc{S}$.

One of our contributions is to show that the condition (\ref{eq:entropy-positiveness}) can be replaced with another mathematical condition which is fully relevant in dynamic games (at least in discrete-time dynamic games). The proposed condition is relevant in any game where an entity (a mediator typically) has to encode the action profiles of the game into messages intended for players having a certain observation structure. Considering that the relevant entity has to encode a sequence of action profiles, the proposed condition consists in assuming that at most one player can change his distribution in an arbitrary manner during the course of the game (single deviations) whereas the other do not change their strategy.
One of the motivations for deriving coding schemes which are resilient \cite{Halpern-podc-2008} to single deviations is therefore to consider the following class of relevant game-theoretic scenarios. This class consists of dynamic games with signals (the game is played several times and players do not observe the actions played perfectly) in which an exogenous entity (see Fig \ref{fig:ReconstructionRobustSide}) sends an additive signal so that players obtain better observations. More motivations for considering this type of scenarios can be found in \cite{LetreustLasaulce(GAMECOMM)11}.

\section{Main technical result}\label{sec:MainTheorem}

\subsection{Definitions and closest existing result}\label{sec:DefLitterature}

In this section, we consider an information source, represented by Fig. \ref{fig:AVCSmodel}, where the symbols are drawn with a distribution which may vary from one stage to another in an arbitrary manner.
Denote by $\mc{S}$ the set of states, $\mc{X}$,  $\mc{Y}$ two sets of symbols and $P(\cdot|s)\in \Delta(\mc{X}\times \mc{Y});s\in \mc{S}$ the distribution of the pair of random variables $(\bd{x},\bd{y})$ when the state is $s\in \mc{S}$; the set $\Delta(\mc{E})$ denotes the set or probability distributions on the generic set $\mc{E}$ that is, the unit simplex associated with $\mc{E}$. For every sequence of states $s^n\in \mc{S}^n$, define the probability distribution of the source sequence $(x^n,y^n)$:
\begin{eqnarray}
\forall s^n\in \mc{S}^n, \ \ P(x^n,y^n|s^n) = \prod_{t=1}^n P(x(t),y(t)|s(t)).
\end{eqnarray}
The arbitrary varying correlated source (AVCS) is defined by the conditional probability $P(\bd{x},\bd{y}|s)$ and by the set of possible sequences $\{s^n\}$ of source states. The component $\bd{x}$ (resp. $\bd{y}$) of the correlated source is received by the first (resp. second) encoder $\mc{C}_1$ (resp. $\mc{C}_2$).
Both encoders want to transmit their source component to the receiver. Our goal is to construct a joint code for which the error probability is upper-bounded in every state sequence $s^n \in \mc{S}^n$.
\begin{definition}
Define an $(n,M_1,M_2)$-code as a triple of functions
\begin{eqnarray}
&f_1& : \mc{X}^n \longrightarrow \mc{M}_1,\\
&f_2& : \mc{Y}^n \longrightarrow \mc{M}_2,\\
&g_2& : \mc{M}_1 \times \mc{M}_2 \longrightarrow \mc{X}^n\times \mc{Y}^n
\end{eqnarray}
where $\mc{M}_i = \{1,2,...,M_i\}$, $i\in \{1,2\}$. With each code is associated an error criterion which has to be minimized.
Define the error probability $P_e^n$ associated to each $(n,M_1,M_2)$-code as follows:
\begin{eqnarray}
P_e^n = \max_{s^n\in \mc{S}^n} P((\bd{x}^n,\bd{y}^n)\neq(\hat{\bd{x}}^n,\hat{\bd{y}}^n)|s^n).
\end{eqnarray}
\end{definition}

\begin{definition}
A rate pair $(R_1,R_2)$ is achievable if for all $\varepsilon>0$, there exists an $(n,M_1,M_2)$-code such that~:
\begin{eqnarray}
\frac{\log M_1}{n} &\leq& R_1+\varepsilon,\\
\frac{\log M_2}{n} &\leq& R_2+\varepsilon,\\
P_e^n &\leq& \varepsilon.
\end{eqnarray}
Denote by $\mc{R}$ the set of achievable rates.
\end{definition}

\begin{theorem}[Ahlswede \cite{Ahlswede(ColoringPart1)79,Ahlswede(ColoringPart2)80}]\label{MainTheorem} Assume that the arbitrary varying correlated source satisfies the entropy positiveness condition~:
\begin{eqnarray}\label{condition:EntropyPositiveness}
\forall s\in \mc{S}, \ \ H(\bd{x}(s)|\bd{y}(s)) \times H(\bd{y}(s)|\bd{x}(s)) >0.
\end{eqnarray}
Then the achievable rate region  $\mc{R}$ equals the following rate region:
\begin{eqnarray*}\label{inequalities:maintheorem}
R_1&\geq & \sup_{P_s\in \Delta(\mc{S})} H(\bd{x}(s)|\bd{y}(s)),\\
R_2&\geq & \sup_{P_s\in \Delta(\mc{S})} H(\bd{y}(s)|\bd{x}(s)),\\
R_1+R_2&\geq & \sup_{P_s\in \Delta(\mc{S})} H(\bd{x}(s),\bd{y}(s)),
\end{eqnarray*}
where the supremum is taken over the probability distributions $P_s\in \Delta(\mc{S})$.
\end{theorem}

This characterization is not valid in general if we remove the entropy positiveness condition.
In the following section, we consider a special but important class of information sources satisfying the
orthogonal deviation condition presented in definition \ref{def:OrthogonalDeviation}.
We will see that this assumption allows us to get rid of the entropy positiveness condition.

\subsection{A new result}

We investigate a class of arbitrary varying sources as described in Fig. \ref{fig:ReconstructionRobustSide} for which
the entropy positiveness condition (\ref{condition:EntropyPositiveness}) is replaced with the orthogonal
deviation condition (\ref{condition:OrthogonalDeviation}). The information source under investigation falls into the class of AVCS because the decoder observes a sequence $y^n$ of side information correlated with $x^n$. Our main result
concerns the source of vector of information $\bd{x}=(\bd{x}_1,\ldots,\bd{x}_K)\in \mc{X}$.

\begin{theorem}\label{theo:main}
Let $P_k(\cdot|s)\in \Delta(\mc{X}_k),\;k\in \mc{K},\;s\in \mc{S}$ the family of source distributions of $\bd{x}_k,\;k\in \mc{K}$ and $\daleth:\Delta(\mc{X}) \longrightarrow \Delta(\mc{Y})$ the transition probability
corresponding to the AVS described by Fig \ref{fig:ReconstructionRobustSide}. The rate~:
\begin{eqnarray}
\mc{R}^{\star} = \max_{k\in \mc{K}} \bigg[\max_{x_k\in \mc{X}_k}H(\bd{x}_{-k}|\bd{y}_{x_k})+\log \chi_k \bigg],\label{eq:theo:main}
\end{eqnarray}
is achievable for the source $\bd{x}=(\bd{x}_1,\ldots,\bd{x}_K)$ and there is no lossless coding scheme providing a better rate.
\end{theorem}
The proof of this theorem is not trivial and a sketch is provided in section \ref{sec:DemoTheoRobustSide}. The purpose of what follows is to explain the notations used in the above theorem and its meaning.

\begin{figure}[!ht]
\begin{center}
\psset{xunit=0.4cm,yunit=0.4cm}
\begin{pspicture}(-4,-1)(16,10)
\psframe(8,4)(9,5)
\psframe(13,4)(14,5)
\pscircle(8.5,2){0.2}
\psdots(0.5,4.5)(4.5,4.5)
\psline[linewidth=1pt]{->}(0.5,1)(0.5,4.2)
\psline[linewidth=1pt]{->}(0.5,8)(0.5,4.8)
\psline[linewidth=1pt]{->}(-3,4.5)(0.2,4.5)
\psline[linewidth=1pt]{->}(-1.97,2.025)(0.3,4.3)
\psline[linewidth=1pt]{->}(-1.97,6.97)(0.3,4.7)
\psline[linewidth=1pt]{->}(0.5,4.5)(8,4.5)
\psline[linewidth=1pt]{->}(9,4.5)(13,4.5)
\psline[linewidth=1pt]{->}(14,4.5)(16,4.5)
\psline[linewidth=1pt](4.5,4.5)(4.5,2)
\psline[linewidth=1pt]{->}(4.5,2)(8,2)
\psline[linewidth=1pt](9,2)(13.5,2)
\psline[linewidth=1pt]{->}(13.5,2)(13.5,4)
\rput[u](0.5,0.5){$\bd{x}_1^n$}
\rput[u](13.5,4.5){$\mc{D}$}
\rput[u](-2.5,2){$\bd{x}_2^n$}
\rput[u](-3.5,4.5){$\bd{x}_3^n$}
\rput[u](-2.5,7.5){$\bd{x}_k^n$}
\rput[u](-1.2,8){$\ldots$}
\rput[u](-3,6.5){$\vdots$}
\rput[u](0.5,8.5){$\bd{x}_K^n$}
\rput[u](8.5,4.5){$\mc{C}$}
\rput[u](8.5,2){$\daleth$}
\rput[u](4.2,5.1){$\bd{x}^n=(\bd{x}_1^n,\ldots,\bd{x}^n_K)$}
\rput[u](15,5){$\hat{\bd{x}}^n$}
\rput[l](13.7,3){$\bd{y}^n$}
\rput[u](11,5){$\bd{m}$}
\end{pspicture}
\caption{The information vector $\bd{x}=(\bd{x}_k)_{k\in \mc{K}}\in \mc{X} = \prod_{k\in \mc{K}}\mc{X}_k$ is drawn with the family of probability distribution $P_k(\cdot|s)\in \Delta(\mc{X}_k),\;k\in \mc{K},\;s\in \mc{S}$ which depend on the state parameter $s\in \mc{S}$. The decoder observes a side information $\bd{y}$ drawn from the transition probability $\daleth(\bd{y}|\bd{x})$. The encoder $\mc{C}$  encodes a sequence of symbols $\bd{x}^n$ and send a message $\bd{m}$ of minimal cardinality $M$ to the decoder $\mc{D}$ such as reconstruct the sequence of symbols $\hat{\bd{x}}^n$ with an error probability arbitrarily small.}\label{fig:ReconstructionRobustSide}
\end{center}
\end{figure}
Consider a family of sets $(\mc{X}_k)_{k\in \mc{K}}$ and a family of distributions $P_k\in \Delta(\mc{X}_k)$ over each component $k\in \mc{K}$.
Denote $\mc{X}=\prod_{k\in \mc{K}}\Delta(\mc{X}_k)$ the product of sets; $P\in \Delta(\mc{X})$ the product probability defined by $P=\bigotimes_{k\in \mc{K}}P_k$ and $P_{-k}\in \Delta(\mc{X}_{-k})$ the probability defined by $P_{-k}=\bigotimes_{j\neq k}P_j$ where the $k$'s component has been removed.

The information source $\bd{x}$ we consider is a vector $\bd{x}=(\bd{x}_1,\ldots,\bd{x}_K)$ of random variables. A sequence of state of the source $s^n\in \mc{S}^n$ is a sequence of probability distribution over one component $j\in \mc{K}$  of the vector $\bd{x}^n=(\bd{x}_1^n,\ldots,\bd{x}_K^n)$.
\begin{definition}\label{def:OrthogonalDeviation}
An arbitrarily varying source of information satisfies the orthogonal deviation condition if the source of information is vectorial $\bd{x}=(\bd{x}_1,\ldots,\bd{x}_K)$ and the set of sates $\mc{S}^n$ is defined by the equation (\ref{condition:OrthogonalDeviation}).
\begin{eqnarray}\label{condition:OrthogonalDeviation}
\mc{S}^n = \cup_{k\in \mc{K}} \Delta(\mc{X}_k)^{\otimes n}
 \end{eqnarray}
 \end{definition}
Consider for example, the sequence $s^n=(Q_j(1),\ldots,Q_j(n))\in \mc{S}^n$ of distributions over the component $j\in \mc{K}$ of the vector $x^n=(x_1^n,\ldots,x_K^n)$ of information. At stage $1\leq t\leq n$ the symbol $x_j(t)$ is drawn according to $Q_j(t)\in \Delta(\mc{X}_j)$ whereas the other symbols $x_k(t)$ are drawn i.i.d. according to the probability distribution $P_k\in \Delta(\mc{X}_k)$. Hence, the vector of information $x(t)$ is drawn, at stage $t\geq 1$, according to the following probability distribution:
\begin{eqnarray*}
P(x_1,\ldots,x_j,\ldots,x_K|s)= [P_1\otimes\ldots\otimes Q_j(t)\otimes \ldots\otimes P_K](x).
\end{eqnarray*}
We consider the situation where the decoder has a side information. It observes, through a transition probability $\daleth : \mc{X} \longrightarrow \Delta(\mc{Y})$, a degraded version $\bd{y}$ of the source symbols $\bd{x}$. The orthogonal deviation condition allow the transition probability $\daleth$ to be deterministic. In such a case, the entropy positiveness condition is not satisfied. Then we conclude that conditions (\ref{condition:EntropyPositiveness}) and (\ref{condition:OrthogonalDeviation}) are disjoint.

\begin{definition}
Define $(n,M)$-code as a triple of functions
\begin{eqnarray}
&f& : \mc{X}^n \longrightarrow \mc{M}, \\
&g& : \mc{M} \times \mc{Y}^n \longrightarrow \mc{X}^n.
\end{eqnarray}
Each code is associated with an error criterion which aims at being minimized.
Define the error probability $P_e^n$ associated to each $(n,M)$-code as follows:
\begin{eqnarray}
P_e^n = \max_{s^n\in \mc{S}^n} P((\bd{x}^n,\bd{y}^n)\neq(\hat{\bd{x}}^n,\hat{\bd{y}}^n)|s^n).
\end{eqnarray}
\end{definition}
\begin{definition}\label{def:AchievableRate}
A rate $R$ is achievable if for all $\varepsilon>0$, there exists a $(n,M)$-code such that:
\begin{eqnarray}
\frac{\log M}{n} &\leq& R+\varepsilon, \\
P_e^n &\leq& \varepsilon.
\end{eqnarray}
Denote $\mc{R}^{\star} $ the infimum of achievable rates $R$.
\end{definition}
For each component $k\in \mc{K}$, define the associated graph $\mc{G}_k$.
\begin{definition}\label{def:Graph}
The graph $\mc{G}_k=(\mc{X}_k,E_k)$ is defined as follows. The source symbols $x_k\in \mc{X}_k$ are the vertex of the graph and there is an edges $e=(x_k,x_k')\in E_k$ between two vertices if and only if:
\begin{eqnarray*}
 & \exists x_{-k}\in  Supp\; P_{-k},\; \exists y\in \mc{Y},\; \exists \delta>0,\text{ s.t. },&\\
 &\min(\daleth(y|x_k,x_{-k}), \daleth(y|x_k',x_{-k}))\geq\delta.&
\end{eqnarray*}
Let $C$ be a set of colors. A coloring of the graph $\mc{G}_k$ is a function $\phi_k :\mc{X}_k \longrightarrow C$ that satisfies:
\begin{eqnarray*}
 e=(x_k,x'_k)\in E_k,\;\Longrightarrow \;\phi_k(x_k)\neq \phi_k(x'_k).
\end{eqnarray*}
\end{definition}
\begin{definition}\label{def:ChromaticNumber}
The chromatic number $\chi_k$ of the graph $\mc{G}_k$ is the cardinality of the minimal coloring of the graph $\mc{G}_k$.
\end{definition}

Denote $\daleth_{x_k}: \mc{X}_{-k}\longrightarrow\Delta(\mc{Y})$ the marginal transition probability of $\daleth$ over the set of signal $\mc{Y}$ when fixing the symbol $x_k\in \mc{X}_k$.
\begin{eqnarray}
\daleth_{x_k} : \mc{X}_{-k} &\longrightarrow& \Delta(\mc{Y}),\\
x_{-k}&\longrightarrow&\daleth_{x_k}(y|x_{-k})=\daleth(y|x_k,x_{-k}).
\end{eqnarray}
 Denote $\bd{y}_{x_k}$ the random signal drawn from the transition $\daleth_{x_k}$. The definitions of $\chi_k$ and $\bd{y}_{x_k}$ are introduced to evaluate the optimal rate $\mc{R}^{\star}$ defined by the equation (\ref{eq:theo:main}).

In the following sections, we provide a sketch of the proof of the theorem \ref{theo:main} (Sec. \ref{sec:DemoTheoRobustSide}) and we apply the derived results to two famous games (Sec. \ref{sec:ApplicationGames}). Note that the theorem \ref{theo:main} holds for mixed strategies and when realizations of the game action profiles are i.i.d. The Shannon theory literature comprises tools \cite{Han(InfoSpectrumBook)06} to extend the theorem \ref{theo:main} to more general setups (e.g., sources with memory or strategies with recall).


\section{Scketch of proof of the theorem \ref{theo:main}}\label{sec:DemoTheoRobustSide}

In this section, we present a sketch of the proof of theorem \ref{theo:main} which will be completely detailed in a separate paper. This section  is divided in two parts called the achievability and the converse.

\subsection{Achievability}
The achievability consists in providing codes that achieves the rate $\mc{R}^{\star}$.
We construct a coding scheme based on graph coloring and statistical tests.
Two points should be considered carefully. First, it is possible that the sequence
of side information $y^n$ provides relevant information to the decoder, for some sequence of state of the source.
Second, the coding scheme of Slepian and Wolf \cite{slepian-it-1973} is not feasible. Indeed, the
transition probability $\daleth$ generating the side information is controlled by
the symbols $\bd{x}_k\in \mc{X}_k$ of each component $k\in \mc{K}$ whose probability
distribution $P_k\in \Delta(\mc{X}_k)$ is not necessarily known.

\textit{The coding scheme.} The encoder performs a statistical
test to determine the component $k\in \mc{K}$ whose sequence of symbols $x_k\in \mc{X}_k$ is the least typical for the distribution $P_k\in \Delta(\mc{X}_k)$.

The sequence of symbols $x_{k} ^n\in \mc{X}_{k}^n $ of the component $k\in \mc{K}$ will be encoded using the coloring of the corresponding graph $\mc{G}_k$. The coloring property of definition \ref{def:Graph} implies that the sequence of side information $y^n$ and the sequence of colors $c^n$ characterize a unique sequence of symbols $x_k^n\in \mc{X}_k^n$.

Other components $x_{-k} ^n\in \mc{X}_{-k}^n $ will be encoded according to the transition probability $\daleth$ and the sequence $x_k^n$.
We construct a partition $(\widetilde{\mc{X}_k},\widetilde{\mc{X}_k^c})$ of the set of symbols $\mc{X}_k$.
If the symbol $x_k \in \mc{X}_k$ is often used (partition $\widetilde{\mc{X}_k}$), the sequence of signals $s_{x_k}^n$, driven by the marginal probability $\daleth_{x_k}$, is long enough to use a source coding with side information of Slepian and Wolf \cite{slepian-it-1973}. Otherwise (partition $\widetilde{\mc{X}_k^c}$), the information $x_{-k} \in \mc{X}_{-k} $ must be encoded directly, without any compression. The result of \cite{slepian-it-1973} states that the sequence of components $x_{-k} \in \mc{X}_{-k} $ can be recovered by the decoder with an arbitrarily small error probability.

\emph{Evaluation of the rate of the code.}
For each $\varepsilon > 0$, there exists a $\bar{n}_1 \in \N$ such that the probability of error of the coding scheme of Slepian and Wolf \cite{slepian-it-1973} is bounded by $\varepsilon > 0$. Let $\bar{n_2}>\frac{\log|\mc{K}|+\bar{n_1}|\mc{X}_k|\log|A_{-k}|}{\varepsilon}$.
We show that for all $n\geq\bar{n_2}  \in \N $, the rate of the code is bounded by the following quantity:

\begin{eqnarray}
\frac{\log M}{n} &=&  \frac{\log \bigg(|\mc{K}|\cdot\chi_k^n\cdot |\mc{X}_{-k}|^{\sum_{x_k\in \widetilde{\mc{X}_k}}n_{x_k}}\bigg)}{n}\nonumber \\
&+& \frac{\log \bigg(\prod_{x_k\in \widetilde{\mc{X}_k^c}}2^{n_{x_k}(H(\bd{x}_{-k}|\bd{y}_{x_k})+2\varepsilon)}\bigg)}{n}\nonumber \\
 &\leq &  \frac{\log|\mc{K}|}{n} + \log\chi_k+ \frac{\bar{n_1}|\widetilde{\mc{X}_k}|}{n}\log|\mc{X}_{-k}|\\
  &+& \sum_{x_k\in \widetilde{\mc{X}_k^c}}\frac{n_{x_k}}{n}(H(\bd{x}_{-k}|\bd{y}_{x_k})+2\varepsilon)\nonumber\\
&\leq&  \max_{x_k\in \mc{X}_k} \bigg[H(\bd{x}_{-k}|\bd{y}_{x_k}) + \log\chi_k\bigg] \\
&+& \frac{\log|\mc{K}|+\bar{n_1}|\mc{X}_k|\log|\mc{X}_{-k}|}{n}+2\varepsilon\nonumber\\
 &\leq& \max_{k\in \mc{K}}\bigg[ \max_{x_k\in \mc{X}_k}H(\bd{x}_{-k}|\bd{y}_{x_k}) + \log\chi_k\bigg] + 3\varepsilon\nonumber \\
&=& \mc{R}^{\star} + 3\varepsilon.
\end{eqnarray}

We provide a sequence of code such that the error probability can be made arbitrarily  small and in the same time the rate of the code is upper bounded by $\mc{R}^{\star}+3\varepsilon$.

\subsection{Converse}

In the converse, we show that no code whose rate is less than  $\mc{R}^{\star}$ is reliable (the error probability is lower bounded away from zero).
To show the converse, we assume that the optimal rate $\mc{R}^{\star}$ is less than the following quantity:
\begin{eqnarray}
 \mc{R}^{\star} < \max_{k\in \mc{K}} \bigg[\max_{x_k\in \mc{X}_k}H(\bd{x}_{-k}|\bd{y}_{x_k})+\log \chi_k  \bigg].
\end{eqnarray}

Suppose the maximum in equation (\ref{eq:theo:main}) is reached for the component $k\in \mc{K}$ and symbol $\bar{x}_k\in \mc{X}_k$. The probability distribution of the vector of actions $x=(x_1,\ldots,x_K)$ is the product of marginal distributions of each component $x_k\in \mc{X}_k$ with $k\in \mc{K}$. The optimum rate $\mc{R}^{\star}$ is then written as the sum of rates of the components $\mc{R}^{\star}=\sum_{k\in \mc{K}} \mc{R}^{\star}_k$. The converse is divided into two parts.

\textit{First,} assume that the encoding scheme has a rate $\mc{R}^{\star}_k<\log\chi_k$. By the property of the minimum coloring, the same color $c_k$ can be generated by two different actions $\tilde{x_k}\in \mc{X}_k$ and $\tilde{x_k}'\in \mc{X}_k$. Then, the probability of error tends to 1 as $ n $ tends to infinity.

\textit{Second,} suppose that the sequence $x_k^n$ is fully reconstructed by the player that decodes but the rate $\mc{R}^{\star}_{-k}$ satisfies:
\begin{eqnarray}
\mc{R}^{\star}_{-k} < H_{s_{\bar{x}_k}}(\bd{x}_{-k}|\bd{y}_{x_k})= \max_{x_k\in \mc{X}_k}H(\bd{x}_{-k}|\bd{y}_{x_k}).
\end{eqnarray}
In this case, the result of Slepian and Wolf \cite{slepian-it-1973} ensures that the probability of error tends to 1 when the sequence of states $s_k^n=\bar{x}_k^n$ and $n$ tends to infinity.

\section{Application to games}\label{sec:ApplicationGames}

The results provided so far can be applied to any discrete-time dynamic game an arbitrary observation structure. We consider the cases of the prisoner's dilemma and the battle of sexes. The information source to be compressed/encoded (by a mediator) is the action profile of the game. The destination (i.e., the decoder) is given by the set of players who observe the played actions through a certain observation structure. So far, we have been assuming that the communication links between the mediator and the players is perfect. Here we assume that each link from the mediator to a given player has a certain capacity denoted by $C_0$.

The strategic form of the prisoner's dilemma is defined by~:
\begin{eqnarray}
G=(\mc{K},(\mc{A}_k)_{k\in \mc{K}}, (u_k)_{k\in \mc{K}})
\end{eqnarray}
where $\mc{K}$ is the set of players $|\mc{K}|=2$, $\mc{A}_k$ is their discrete action sets $\mc{A}_1=\{T,B\}$ and $\mc{A}_2=\{L,R\}$ whereas $u_k$ is the utility function of player $k\in \mc{K}$ such that :
\begin{eqnarray}
u_k  &:& \mc{A}_1 \times \mc{A}_2 \longrightarrow \R, \quad k\in \mc{K}.
\end{eqnarray}
For the prisoner's dilemma, the utility function writes as the tale of Fig. \ref{fig:PayoffPrisonerDilemma}.
When player 1  plays $B$ and player 2  plays $L$, the utility of player 1 is 4 and the one of player 2 is 0.
\begin{figure}[!ht]
\begin{center}
\psset{xunit=1cm,yunit=0.5cm}
\begin{pspicture}(-1,-1)(5,5)
\psframe(0,0)(4,4)
\psline(0,2)(4,2)
\psline(2,0)(2,4)
\rput(1,3){$3,3$}
\rput(3,3){$0,4$}
\rput(1,1){$4,0$}
\rput(3,1){$1,1$}
\rput(-0.5,3){$T$}
\rput(-0.5,1){$B$}
\rput(1,5){$L$}
\rput(3,5){$R$}
\end{pspicture}
\caption{The utility matrix of the prisoner's dilemma.}\label{fig:PayoffPrisonerDilemma}
\end{center}
\end{figure}

Suppose that the player play according to a probability distribution $P\in \Delta(\mc{A}_1)\times \Delta( \mc{A}_2)$ (mixed strategies). The problem is to evaluate the rate associated with the information source (generating the action profiles) knowing that one of the players is potentially deviating using a different probability distribution. Also suppose that the destination of the communication system has a degraded version $s^n$ of the source sequence $a^n$ from a transition probability:
\begin{eqnarray}
\daleth : \mc{A} \longrightarrow \Delta(\mc{S}),
\end{eqnarray}
where $\mc{A}= \mc{A}_1 \times \mc{A}_2$.

\textbf{Numerical results.} In order to illustrate the above result, we need to fix a transition probability $\daleth$ from the set of actions to a set of signals $\{s_1,s_2,s_3,s_4\}$.
 \begin{figure}[h!]
\begin{center}
\psset{xunit=0.5cm,yunit=0.5cm}
\begin{pspicture}(-2,1)(14.5,7)
\psframe(-2,2)(-1,3)
\psframe(-2,6)(-1,7)
\psframe(4,4)(5,5)
\psframe(10,2)(11,3)
\psframe(10,6)(11,7)
\pscircle(4.5,1.5){0.25}
\pscircle(7.5,4.5){0.25}
\psdots(0.5,4.5)(2.5,4.5)(8.5,4.5)(8.5,1.5)
\psline[linewidth=1pt]{->}(-1,3)(0.3,4.3)
\psline[linewidth=1pt]{->}(-1,6)(0.3,4.7)
\psline[linewidth=1pt]{->}(0.5,4.5)(4,4.5)
\psline[linewidth=1pt]{->}(5,4.5)(7,4.5)
\psline[linewidth=1pt](8,4.5)(8.5,4.5)
\psline[linewidth=1pt]{->}(8.5,4.5)(10,3)
\psline[linewidth=1pt]{->}(8.5,4.5)(10,6)
\psline[linewidth=1pt]{->}(11,2.5)(14.5,2.5)
\psline[linewidth=1pt]{->}(11,6.5)(14.5,6.5)
\psline[linewidth=1pt](2.5,4.5)(2.5,1.5)
\psline[linewidth=1pt]{->}(2.5,1.5)(4,1.5)
\psline[linewidth=1pt](5,1.5)(10.5,1.5)
\psline[linewidth=1pt]{->}(10.5,1.5)(10.5,2)
\psline[linewidth=1pt](8.5,1.5)(8.5,3.5)
\psline[linewidth=1pt](8.5,3.5)(9.35,3.5)
\psarc[linecolor=black,linewidth=1pt](9.5,3.5){0.09375}{0}{180}
\psline[linewidth=1pt](9.65,3.5)(10.5,3.5)
\psline[linewidth=1pt]{->}(10.5,3.5)(10.5,6)
\rput[u](-1.5,2.5){$J_2$}
\rput[u](-1.5,6.5){$J_1$}
\rput[u](10.5,2.5){$J_2$}
\rput[u](10.5,6.5){$J_1$}
\rput[u](4.5,4.5){$\C$}
\rput[u](4.5,1.5){$\daleth$}
\rput[u](7.5,4.5){$\daleth_0$}
\rput[u](-0.5,2.5){$\bd{a}_2^n$}
\rput[u](-0.5,6.5){$\bd{a}_1^n$}
\rput[u](2.2,5.45){$\bd{a}^n=(\bd{a}_1^n,\bd{a}^n_2)$}
\rput[u](13.2,7){$\hat{\bd{a}}^n=(\hat{\bd{a}}_1^n,\hat{\bd{a}}^n_2)$}
\rput[u](13.2,3){$\hat{\bd{a}}^n=(\hat{\bd{a}}_1^n,\hat{\bd{a}}^n_2)$}
\rput[l](10.7,1.6){$\bd{s}^n$}
\rput[l](10.7,5.6){$\bd{s}^n$}
\rput[u](6,5){$\bd{v}^n$}
\rput[u](9.5,4.5){${\bd{z}^n}$}
\end{pspicture}
\caption{Players $J_1$ and $J_2$ observe, through $\daleth$, a noisy version $\bd{s}^n$ of the sequence of actions $\bd{a}^n=(\bd{a}_1^n,\ldots,\bd{a}^n_K)$. The encoder $\C$ transmits, through the channel $\daleth_0$, the sequence of encoded actions $\bd{a}^n$ to the players $J_1$ and $J_2$. The players recover the sequence of past actions $\hat{\bd{a}}^n={\bd{a}}^n$ with an error probability arbitrarily small if and only if the rate $\mc{R}^{\star}$ of the source is less than the capacity $C_0$ of the channel $\daleth_0$, fixed to $1.7$ bit per second.}\label{fig:ReconstructionRobustSideCapa}
\end{center}
\end{figure}
The transition probability is defined as follows. When the action $(T,L)$ is played, the signal of the decoder is drawn from the probability distribution described by Fig. \ref{fig:TransitionProbability}. More precisely, the probability of observing the right signal $s_1$ (which corresponds to the actions $(T,L)$) equals $1-\frac{3}{4}\varepsilon$ whereas the probability of observing another signal ($s_2$, $s_3$ or $s_4$) is equal to $\frac{1}{4}\varepsilon$. The parameter $\varepsilon\in [0,1]$ allows us to compare both situations: when $\varepsilon =0$, the actions $a=(a_1,a_2) \in \mc{A}$ are perfectly recovered and when $\varepsilon =1$, the signal do not provide any relevant information.
When the action profile $(T,R)$ (resp. $(B,L)$ or $(B,R)$) is played the corresponding signal $s_2$ (resp. $s_3$ or $s_4$) is also drawn with probability $1-\frac{3}{4}\varepsilon$ whereas the other signals are drawn with probability $\frac{1}{4}\varepsilon$.

\begin{figure}[!ht]
\begin{center}
\psset{xunit=1cm,yunit=0.5cm}
\begin{pspicture}(-1,-1)(5,5)
\psframe(0,0)(4,4)
\psline(0,2)(4,2)
\psline(2,0)(2,4)
\rput(1,3){$1-\frac{3}{4}\varepsilon$}
\rput(3,3){$\frac{1}{4}\varepsilon$}
\rput(1,1){$\frac{1}{4}\varepsilon$}
\rput(3,1){$\frac{1}{4}\varepsilon$}
\rput(1.8,2.3){$s_1$}
\rput(3.8,2.3){$s_2$}
\rput(1.8,0.3){$s_3$}
\rput(3.8,0.3){$s_4$}
\end{pspicture}
\caption{The transition probability $\daleth$ of signals received by the decoder $\mc{D}$ when the joint action $(T,L)$ is played.}\label{fig:TransitionProbability}
\end{center}
\end{figure}

\begin{figure}[!ht]
\begin{center}
\psset{xunit=1cm,yunit=0.5cm}
\begin{pspicture}(-1,-1)(5,1)
\psdots(0,0)(2,0) (4,0)(6,0)
\psline(0,0)(2,0)
\psline(4,0)(6,0)
\rput[u](0,0.4){$T$}
\rput[u](2,0.4){$B$}
\rput[u](4,0.4){$L$}
\rput[u](6,0.4){$R$}
\rput(1,-0.8){$\mc{G}_1$}
\rput(5,-0.8){$\mc{G}_2$}
\end{pspicture}
\caption{The graphs $\mc{G}_1$ and $\mc{G}_2$ of the players for $\varepsilon>0$. When  $\varepsilon = 0$, there is no edges in both graphs.}\label{fig:GraphPlayers}
\end{center}
\end{figure}
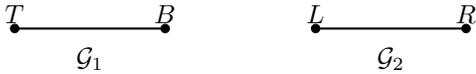

Figure \ref{fig:GraphPlayers} represents the graphs $\mc{G}_1$ and $\mc{G}_2$ over player's action which correspond to the above transition probability $\daleth$ and the definition \ref{def:Graph} of Sec. \ref{sec:MainTheorem}.

As soon as $\varepsilon>0$ the graph the players are described in Fig.
\ref{fig:GraphPlayers}. Note that if $\varepsilon>0$ the coloring of both graphs is of cardinality of 2 and thus $\log \chi_1=\log \chi_2=1$.
If $\varepsilon=0$ the pair of graphs have no edges.

We evaluate the rate of such a source in order to transmit the sequence of actions. However, the transmission of the source of information over a channel is not always feasible. A natural question arises and is illustrated by figure \ref{fig:ReconstructionRobustSideCapa}. \textit{Is it possible to transmit the source of action profiles through a channel with fixed capacity $C_0$ ?}
Recall that the source of information $\bd{a}$ with side information $\bd{s}$ can be transmitted reliably over a channel $\daleth_0$ with capacity $C_0$, if $\mc{R}^{\star}<C_0$.
We evaluate the optimal compression rate thanks to theorem \ref{theo:main}.
\begin{eqnarray}
\mc{R}^{\star} = \max_{k\in \mc{K}} \bigg[\max_{a_k\in \mc{A}_k}H(\bd{a}_{-k}|\bd{s}_{a_k})+\log \chi_k\bigg].
\end{eqnarray}

\begin{figure}
\centering
\includegraphics[width=0.50\textwidth]{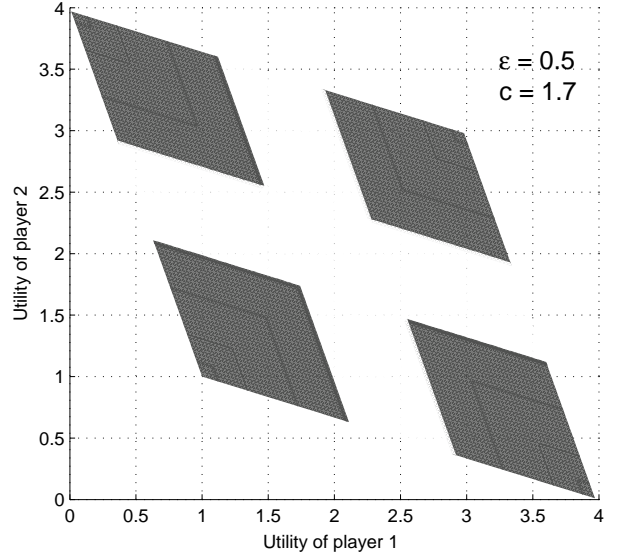}
\caption{Achievable utility region for the prisoner's dilemma under capacity constraints.}\label{fig:prisoner'sDilemma}
\end{figure}

Suppose that the capacity of channel $\daleth_0$ equals $C_0=1.7$ (in bit/symbol or bit/sample)and the parameter $\varepsilon=0.5$, we will see that only a subset of the feasible utility region can be achieved, for the communication system of Fig. \ref{fig:ReconstructionRobustSideCapa}, by an i.i.d. strategy  (see e.g., Fig. \ref{fig:prisoner'sDilemma} for the game of the prisoner's dilemma). For each pair of mixed actions $P_1\in \Delta(\mc{A}_1)$ and $P_2\in \Delta(\mc{A}_2)$, we evaluate the rate $ \mc{R}^{\star} $ of encoding the corresponding sequence of actions.
If it satisfies the channel capacity condition $\mc{R}^{\star}< C_0$, the sequence of actions can be recovered even if one of the players deviates from $P_k \in \Delta(\mc{A}_k)$.
The Fig. \ref{fig:prisoner'sDilemma} represents the region of utility vectors whose corresponding mixed actions satisfy the channel condition $\mc{R}^{\star}< C_0$.

\begin{figure}
\centering
\includegraphics[width=0.50\textwidth]{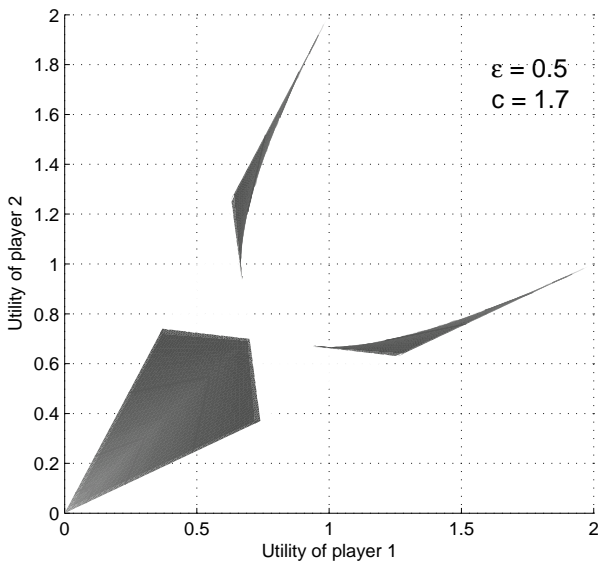}
\caption{Achievable utility region for the battle of sexes under capacity constraints.}
\label{fig:BattleofSexes}
\end{figure}

We consider also the  game of the "battle of sexes" whose utility matrices are represented in Fig. \ref{fig:BattleofSexes}.
For the channel capacity of $C_0=1.7$ bit/symbol and the same transition probability with precision parameter $\varepsilon=0.5$ (see Fig. \ref{fig:TransitionProbability}), we obtain the  achievable utility region of  Fig. \ref{fig:BattleofSexes}.

Remark that in both examples, we consider only mixed strategies. The problem of considering correlated strategies is much more complicated.
Indeed, we need to define properly an arbitrary varying source where the actions and  the deviations may be correlated. It would be also possible to achieve the same utility vector with various probability distributions. In that case, the optimal admissible rates may belong to an interval. Depending on the formulation of the problem, the capacity constraints may be defined with respect to the lowest or the largest optimal rate associated with a utility vector.
In game theory, most of the player's strategy are not generated i.i.d. It should be of interest to extend our results to the case of a information source compatible with the game theoretical strategies using information spectrum methods \cite{Han(InfoSpectrumBook)06}.

\begin{figure}[!ht]
\begin{center}
\psset{xunit=1cm,yunit=0.5cm}
\begin{pspicture}(-1,-1)(5,5)
\psframe(0,0)(4,4)
\psline(0,2)(4,2)
\psline(2,0)(2,4)
\rput(1,3){$0,0$}
\rput(3,3){$2,1$}
\rput(1,1){$1,2$}
\rput(3,1){$0,0$}
\rput(-0.5,3){$T$}
\rput(-0.5,1){$B$}
\rput(1,5){$L$}
\rput(3,5){$R$}
\end{pspicture}
\caption{The utility matrix of the battle of sexes.}\label{fig:PayoffBattleSexes}
\end{center}
\end{figure}

\section{Conclusion}\label{sec:conclusion}

We have investigate the problem of encoding arbitrary varying correlated sources also considered by
Ahlswede in 1980 \cite{Ahlswede(ColoringPart1)79,Ahlswede(ColoringPart2)80}. The ``entropy
positiveness condition'' (\ref{condition:EntropyPositiveness}) is replaced by an ``orthogonal
deviation condition'' (\ref{condition:OrthogonalDeviation}) over the states of the source.
Both conditions are disjoint because there exists some information sources that
satisfies one condition without satisfying the other one. The main result characterizes the
optimal rate of the source of information and a sketch of the proof is presented.
We provide a game-theoretical application of our main result~: a characterization
of the rate of a source of player's actions in a long run game. Our coding scheme is resilient to
any unilateral deviation of the player's strategies. In conclusion, the coding scheme cannot be
manipulated along the duration of the game. The coloring techniques of Ahlswede are directly
involved in the characterization of the optimal rate of an information source.

\section*{Acknowledgment}\label{acknoledgment}

The authors would like to thank Cagatay Dikici for constructive comments.

\bibliographystyle{plain}
\bibliography{BiblioMael}

\begin{thebibliography}{10}

\bibitem{Ahlswede(ColoringPart1)79}
R.~Ahlswede.
\newblock Coloring hypergraphs: A new approach to multi-user source coding,
  part 1.
\newblock {\em Journal of combinatorics, information and system sciences},
  4(1):76--115, 1979.

\bibitem{Ahlswede(ColoringPart2)80}
R.~Ahlswede.
\newblock Coloring hypergraphs: A new approach to multi-user source coding,
  part 2.
\newblock {\em Journal of combinatorics, information and system sciences},
  5(3):220--268, 1980.

\bibitem{Berger(Game)71}
T.~Berger.
\newblock The source coding game.
\newblock {\em IEEE Trans. on Information Theory}, IT-17(1):71--76, Jan. 1971.

\bibitem{Dobrusin(Indiv)63}
R.L. Dobru\v{s}in.
\newblock Individual methods for transmission of information for discrete
  channels without memory and messages with independent components.
\newblock {\em Sov. Math.}, 4:253--256, 1963.

\bibitem{Dobrusin(Unif)63}
R.L. Dobru\v{s}in.
\newblock Unified methods of optimal quantizing of messages.
\newblock {\em Sov. Math.}, 4:284--292, 1963.

\bibitem{Gallager(AVS)76}
R.~G. Gallager.
\newblock Source coding with side information and universal coding.
\newblock In {\em IEEE International Symposium on Information Theory}, Renneby,
  Sweden, 1976.

\bibitem{Halpern-podc-2008}
J.~Halpern.
\newblock Beyond {N}ash equilibrium: Solution concepts for the 21st century.
\newblock In {\em ACM Proc. of PODC'08}, Toronto, Canada, August 2008.

\bibitem{Han(InfoSpectrumBook)06}
T.~S. Han.
\newblock {\em Information-spectrum methods in information theory}.
\newblock Springer, 2003.

\bibitem{LetreustLasaulce(GAMECOMM)11}
M.~LeTreust and S.~Lasaulce.
\newblock The price of re-establishing almost perfect monitoring in games with
  arbitrary monitoring structures.
\newblock {\em ACM Proc. of the 4th International Workshop on Game Theory in
  Communication Networks (GAMECOMM11), Cachan (Paris), France}, 2011.

\bibitem{shannon-bell-1948}
C.~E. Shannon.
\newblock A mathematical theory of communication.
\newblock {\em Bell System Technical Journal}, 27:379--423, 1948.

\bibitem{slepian-it-1973}
D.~Slepian and J.~K. Wolf.
\newblock Noiseless coding of correlated information sources.
\newblock {\em IEEE Trans. on Information Theory}, 19:471--480, July. 1973.

\end{thebibliography}

\end{document}